\documentclass[twocolumn]{aastex631}

\usepackage{graphicx}
\usepackage{amsmath}

\newcommand{\teff}{\ensuremath{\mathrm{T_{eff}}}}

\newcommand{\logg}{\ensuremath{\mathrm{\log g}}}

\DeclareRobustCommand{\ion}[2]{\textup{#1\,\textsc{\lowercase{#2}}}}

\begin{document}

\title{On the Pair-Instability Supernova origin of J1010+2358\thanks{Based on observations made with ESO VLT at the La Silla Paranal observatory under program ID 112.26WJ.001}}
\shorttitle{PISN}
\shortauthors{Skúladóttir et al.}

\newcommand{\red}{\textcolor{red}}

\author[0000-0001-9155-9018]{Ása Skúladóttir}
\affiliation{Dipartimento di Fisica e Astronomia, Universit\'a degli Studi di Firenze, Via G. Sansone 1, I-50019 Sesto Fiorentino, Italy.}

\author[0000-0002-3524-7172]{Ioanna Koutsouridou}
\affiliation{Dipartimento di Fisica e Astronomia, Universit\'a degli Studi di Firenze, Via G. Sansone 1, I-50019 Sesto Fiorentino, Italy.}

\author[0000-0001-9647-0493]{Irene Vanni}
\affiliation{Dipartimento di Fisica e Astronomia, Universit\'a degli Studi di Firenze, Via G. Sansone 1, I-50019 Sesto Fiorentino, Italy.}

\author[0000-0002-3181-3413]{Anish M. Amarsi}
\affiliation{Theoretical Astrophysics, Department of Physics and Astronomy, Uppsala University, Box 516, SE-751 20 Uppsala, Sweden}

\author[0000-0002-9750-1922]{Romain Lucchesi}
\affiliation{Dipartimento di Fisica e Astronomia, Universit\'a degli Studi di Firenze, Via G. Sansone 1, I-50019 Sesto Fiorentino, Italy.}

\author[0000-0001-7298-2478]{Stefania Salvadori}
\affiliation{Dipartimento di Fisica e Astronomia, Universit\'a degli Studi di Firenze, Via G. Sansone 1, I-50019 Sesto Fiorentino, Italy.}

\author[0000-0001-5200-3973]{David S. Aguado}
\affiliation{Instituto de Astrofísica de Canarias, Vía Láctea, 38205 La Laguna, Tenerife, Spain}
\affiliation{Universidad de La Laguna, Departamento de Astrof\'{\i}sica,  38206 La Laguna, Tenerife, Spain.}

\begin{abstract}{

The first (Pop~III) stars formed only out of H and He and were likely more massive than present-day stars. Massive Pop~III stars in the range $\rm140-260\,M_\odot$ are predicted to end their lives as pair-instability supernovae (PISNe), enriching the environment with a unique abundance pattern, with high ratios of odd to even elements. Recently, the most promising candidate for a pure descendant of a zero-metallicity massive PISN (260\,M$_{\odot}$) was discovered by the LAMOST survey, the star J1010+2358. However, key elements to verify the high PISN contribution, C and Al, were missing from the analysis. To rectify this, we obtained and analyzed a high-resolution VLT/UVES spectrum, correcting for 3D and/or non-LTE effects. Our measurements of both C and Al give much higher values ($\sim1$\,dex) than expected from a 260\,M$_{\odot}$ PISN. Furthermore, we find significant discrepancies with the previous analysis, and therefore a much less pronounced odd-even pattern. Our results show that J1010+2358 cannot be a pure descendant of a 260\,M$_{\odot}$ PISN. Instead, we find that the best fit model consists of a 13\,M$_{\odot}$ Pop~II core-collapse supernova combined with a Pop~III supernova. Alternative, less favored solutions ($\chi^2/\chi^2_{\rm best}\approx2.3$) include a 50\% contribution from a 260\,M$_{\odot}$ PISN, or a 40\% contribution from a Pop~III type Ia supernova. Ultimately, J1010+2358 is certainly a unique star giving insights into the earliest chemical enrichment, however, this star is not a pure PISN descendant.

}

\end{abstract}

\section{Introduction} \label{sec:intro}

When trying to understand the first (Pop~III) stars in the Universe, constraining their initial mass function (IMF) is of fundamental importance. The IMF dictates the amount of ionizing radiation and feedback that the first stellar generation produced, as well as the amount and composition of metals they distributed to the environment. 
Although the Pop~III IMF is still unknown, the general consensus is that Pop~III stars were likely more massive than stars born today, possibly up to a thousand times the mass of the sun \citep[e.g.][]{Hirano15,Sharda2020}. 
If the Pop~III IMF was indeed top-heavy, the very massive first stars, $140\leq M_{\star}/\text{M}_{\odot}
\leq 260$, should have ended their lives as pair-instability supernovae (PISNe). 
Their chemical signature is predicted to be characterised by a strong odd-even effect, i.e.~high abundances of even-Z elements relative to their odd-Z neighbours \citep{Heger02,Takahashi18}.

In the previous decade, several studies have proposed candidates for PISN desendants, i.e.~stars that obtained a large fraction ($\geq50\%$) of their metals from a PISN progenitor \citep{Aoki14,Salvadori19,Aguado23}. Recently, the most promising candidate was discovered by the LAMOST survey, \citep{Xing23}. In their initial analysis, the star, J1010+2358, was found to have $\rm[Fe/H]=-2.4$, consistent with predictions for PISN descendants (e.g.~\citealt{Salvadori19}), and it also showed a strong odd-even effect in its abundance pattern. Thus, \citet{Xing23} proposed J1010+2358 to be a pure descendant of a 260\,M$_{\odot}$ PISN. These remarkable results are some of the most promising observational evidence for the existence of PISN, and massive Pop~III stars in general.

By using the discovery of this single PISN descendant, J1010+2358, \citet{Koutsouridou24} were able to make the tightest constraints ever proposed for the Pop~III IMF, under the assumption that all of its metals were inherited from a massive PISN (260\,M$_\odot$). However, \citet{Koutsouridou24} also found that according to the abundance pattern provided by \citet{Xing23}, the fraction of the star's metals coming from PISN could be anywhere in the range $10-100\%$. Furthermore, they were only able to use J1010+2358 to put constraints on the Pop~III IMF if the star received $\geq70\%$ of its metals from a PISN. \citet{Koutsouridou24} found that C would be crucial for distinguishing between different levels of PISN contribution; but this element was not included in the analysis of \citet{Xing23}. In addition, \citet{Jeena24} found another possible solution for J1010+2358, as models of Pop~III $12-14$\,M$_\odot$ core-collapse SNe (ccSNe), with negligible fallback after the explosion, were able to provide an equally good fit as that of a 260\,M$_\odot$ PISN. Several elemental abundances can discriminate between a PISN or a ccSN progenitor, out of which C and Al are the most feasible from an observational standpoint. 

With the aim of distinguishing between these different scenarios, we obtained high-quality ESO VLT/UVES spectrum to measure both C and Al in this star. We will therefore be able to verify or reject J1010+2358 as the most promising observational evidence for the existence of massive zero-metallicity PISNe.

\section{Data analysis} \label{sec:style}
\subsection{Observations and data reduction}

The star J1010+2358 from \citet{Xing23} is located at RA=10:10:51.9, Dec.=+23:58:50.164, with magnitudes: $G=15.8$, $G_{\text{BP}}=16.1$, and $G_{\text{RP}}=15.3$. The star was followed-up in January 2024 with the ESO VLT/UVES spectrograph for a total of $9\times3000$\,s in exposure time. The average airmass was 1.6, and the average seeing 0.6''. The blue arm was centered on 437\,nm (376$-$498\,nm) with a slit width of 0.8'', resulting in a resolution of $\rm R_{\text{blue}}\approx55\,000$, and $\rm S/N=50$\,pix$^{-1}$ at 432\,nm. Our S/N is thus comparable to that of the Subaru spectra presented in \citet{Xing23}, but here the resolution is higher ($\rm R_{\text{Xing}}=36\,000$). The red arm was centered on 760\,nm (569$-$927\,nm) with a slit width of 1.0'', a resolution of $\rm R_{\text{red}}\approx42\,000$, S/N=90\,pix$^{-1}$ at 586\,nm, and S/N=100\,pix$^{-1}$ at 900\,nm.

The line-of-sight velocity of the star was measured from the more line-rich, higher-resolution blue UVES spectrum to be $v_{\text{los}}=-99.4\pm1.0$\,km/s, where \citet{Xing23} measured $v_{\text{los}}^{\text{Xing}}=-101.8\pm0.7$\,km/s. Since neither study made an effort in correcting for possible offsets in their respective spectrographs, and the difference between those measurements is relatively small, $\Delta{v_{los}}=2.4\pm1.2$\,km/s, we refrain from drawing any conclusion regarding binarity.

\subsection{Stellar atmosphere models and linelists}

The adopted stellar atmosphere models are from MARCS\footnote{\url {https://marcs.astro.uu.se/}} \citep{Gustafsson08} for stars with a standard composition, 1D, and assuming local thermodynamic equilibrium (LTE), interpolated to match the stellar parameters of J1010+2358. The abundance analysis was carried out with the spectral synthesis code TURBOSPEC\footnote{\url{https://ascl.net/1205.004}} \citep{AlvarezPlez98,Plez12}. Atomic data were adopted from the VALD\footnote{\url{http://vald.astro.uu.se}} database (\citealt{Kupka99}), while the molecular list for CH was obtained from \citet{Masseron14}.

\subsection{Atmopsheric parameters}

The effective temperature, \teff, was determined using Gaia DR3 photometry and the calibration of \citet{Mucciarelli21}. We adopt $\teff(G_{\text{BP}}-G_{\text{RP}})=5839\pm83$\,K, where the error comes from the $\sigma$ in the calibration, while the uncertainty arising from errors in [Fe/H] and the photometry is negligible. We note that the average \teff{} based on all three Gaia colors $(G_{\text{BP}}-G_{\text{RP}})$, $(G_{\text{BP}}-G)$, and $(G_{\text{BP}}-G_{\text{RP}})$ also gives $\teff(\text{avg})=5839$\,K, with $\sigma=28$\,K. This is in excellent agreement with \citet{Xing23}, who determined $\teff(\text{Xing, spec})=5860\pm120$\,K based on spectroscopy, and also reported a photometric $\teff(\text{Xing, phot})=5810$\,K, from the $(V-K)_0$ color.

The microturbulence, $v_\textsl{t}$, was inferred by requiring that the [Fe/H] from \ion{Fe}{I} lines does not show any trend with reduced equivalent width, $\rm \log(EW/\lambda)$.  In this way we found $v_\textsl{t}=1.37\pm0.20$\,km/s, in good agreement with the value $v_\textsl{t}^{\text{Xing}}=1.50\pm0.25$\,km/s from \citet{Xing23}.

Our analysis of the surface gravity, $\log{g}$, places the star squarely on the main-sequence, rather than on the sub-giant branch as found by \citet{Xing23}.
We measured $\log{g}=4.72\pm0.13$ by using the Gaia eDR3 distance to this star, $r_{pgeo}=1066.6^{+40.6}_{-48.4}$\,pc \citep{Bailer-Jones21}, which allows us to estimate the photometric $\log{g}$ via the standard relation:
\begin{equation}
\small
\log g_{\star}=\log g_{\odot}+\log{\frac{\text{M}_{\star}}{\text{M}_{\odot}}}+  4\log{ \frac{\teff_{,\star}}{\teff_{,\odot}} }+0.4\,(\text{M}_{\text{bol,}\star}-\text{M}_{\text{bol,}\odot})
\end{equation}
where we assume $\text{M}_\star~=~0.8\pm0.2 $~M$_\odot$. The solar values used are the following: $\log~g_\odot=4.44$, $\teff_{,\odot}=5772$\,K and $\text{M}_{\text{bol,}\odot}=4.74$. 
This Gaia-based result is consistent with the isochrone-fitted value of $\log{g}_{\text{iso}}=4.59$, from the Yonsei-Yale isochrones \citep{Demarque04} for a 13~Gyr old star  with $\rm[Fe/H]=-2.4$ and $\teff= 5839$\,K (Fig.~\ref{fig:iso}).
However, our $\log{g}$ result is in $4\sigma$ tension with the value of $\log{g}=3.6\pm0.2$ inferred by \citet{Xing23} from spectroscopy. We note that by adopting this lower $\log{g}$, we were not able to maintain ionization balance, $\rm [\ion{Fe}{I}/{Fe}{II}]_{\rm LTE}\approx+0.3$, in LTE. Taking into account departures from LTE (NLTE effects) and errors from using one dimensional (1D) model atmospheres should furthermore act to make this ionisation disequilibrium more severe \citep[e.g.][]{Amarsi16,Amarsi22}. Therefore, we adopt the Gaia-based result, $\log{g}=4.72\pm0.13$.

\begin{table}
\caption{Chemical abundances of J1010+2358, including the elemental species (El. Sp.), number of lines, $N_l$, NLTE corrections, $\rm\Delta_{\text{NLTE}}=[X/H]_{\text{NLTE}}-[X/H]_{\text{LTE}}$, and error, $\delta_{\text{[X/Fe]}}$. We adopt $\rm [Fe/H]=[\ion{Fe}{II}/H]$ (Sec.~\ref{sec:iron}).
}
\label{tab:abu}
\normalsize
\tabcolsep=0.06cm
\renewcommand{\arraystretch}{1.1}

\centering
\footnotesize
\begin{tabular}{lccrrrrc}
\hline\hline
El. Sp.	&	$\log\epsilon_{\odot}$	&	$N_l$	&	$\log\epsilon$	&	[X/Fe]	&	$\Delta_{\text{NLTE}}$	&	$\rm [X/Fe]_{\text{NLTE}}$	&	$\delta_{\text{[X/Fe]}}$	\\
\hline																
\ion{Li}{I}	&	1.05	&	1	&$	1.95	$&$	3.43	$&$	-0.05	$&$	3.25	$&	0.10	\\
CH$^{a}$	&	8.46	&	$-$	&$	6.35	$&$	0.42	$&$	-0.35	$&$	-0.06	$&	0.14	\\
\ion{C}{I}$^{b}$	&	8.46	&	1	&$	5.91	$&$	-0.02	$&$	-0.07	$&$	-0.22	$&	0.28	\\
N (CN)	&	7.83	&	$-$	&$	<6.55	$&$	<1.25	$&$	0.00	$&$	<1.12	$&	$-$	\\
\ion{O}{I}	&	8.69	&	1	&$	<6.85	$&$	<0.69	$&$	-0.03	$&$	<0.53	$&	$-$	\\
\ion{Na}{I}	&	6.22	&	2	&$	2.43	$&$	-1.26	$&$	-0.03	$&$	-1.42	$&	0.11	\\
\ion{Mg}{I}$^{b}$	&	7.55	&	3	&$	4.27	$&$	-0.75	$&$	0.10	$&$	-0.78	$&	0.13	\\
\ion{Al}{I}	&	6.43	&	2	&$	2.30	$&$	-1.60	$&$	0.40	$&$	-1.33	$&	0.13	\\
\ion{Si}{I}	&	7.51	&	1	&$	4.40	$&$	-0.58	$&$	0.02	$&$	-0.69	$&	0.18	\\
\ion{K}{I}	&	5.07	&	1	&$	<2.60	$&$	<0.06	$&$	-0.12	$&$	<-0.19	$&	$-$	\\
\ion{Ca}{I}	&	6.30	&	4	&$	3.34	$&$	-0.43	$&$	0.07	$&$	-0.49	$&	0.10	\\
\ion{Sc}{II}	&	3.14	&	1	&$	-0.09	$&$	-0.70	$&$	0.10	$&$	-0.73	$&	0.11	\\
\ion{Ti}{II}	&	4.97	&	11	&$	1.97	$&$	-0.47	$&$	0.02	$&$	-0.58	$&	0.04	\\
\ion{V}{I}	&	3.90	&	2	&$	<1.30	$&$	<-0.07	$&$	0.25	$&$	<0.05	$&	$-$	\\
\ion{Cr}{I}	&	5.62	&	7	&$	2.87	$&$	-0.22	$&$	0.19	$&$	-0.16	$&	0.08	\\
\ion{Mn}{I}	&	5.42	&	4	&$	2.29	$&$	-0.60	$&$	0.09	$&$	-0.64	$&	0.12	\\
\ion{Fe}{I}$^{b}$	&	7.46	&	179	&$	4.92	$&$	-0.01	$&$	0.24	$&$	0.10	$&	0.07	\\
\ion{Fe}{II}$^{b,c}$	&	7.46	&	6	&$	4.93	$&$	-2.53	$&$	0.13	$&$	-2.40	$&	0.06	\\
\ion{Co}{I}	&	4.94	&	6	&$	2.23	$&$	-0.18	$&$	0.56	$&$	0.25	$&	0.09	\\
\ion{Ni}{I}	&	6.20	&	5	&$	3.40	$&$	-0.27	$&$	0.24	$&$	-0.16	$&	0.10	\\
\ion{Zn}{I}	&	4.56	&	1	&$	<2.00	$&$	<-0.03	$&$	0.05	$&$	<-0.11	$&	$-$	\\
\ion{Sr}{II}	&	2.83	&	2	&$	-1.18	$&$	-1.48	$&$	0.15	$&$	-1.46	$&	0.08	\\
\ion{Ba}{II}	&	2.27	&	1	&$	<-1.75	$&$	<-1.49	$&$	0.11	$&$	<-1.51	$&	$-$	\\
\hline \hline 
\multicolumn{8}{l}{\fontsize{6.1}{4}\selectfont $^{a)}$ 3D LTE.}\\
\multicolumn{8}{l}{\fontsize{6.1}{4}\selectfont $^{b)}$ 3D NLTE.}\\
\multicolumn{8}{l}{\fontsize{6.1}{4}\selectfont $^{c)}$ [Fe/H] is listed instead of [X/Fe].}\\							

\end{tabular}

\end{table}

\begin{figure*}
\centering
\includegraphics[width=0.59\linewidth]{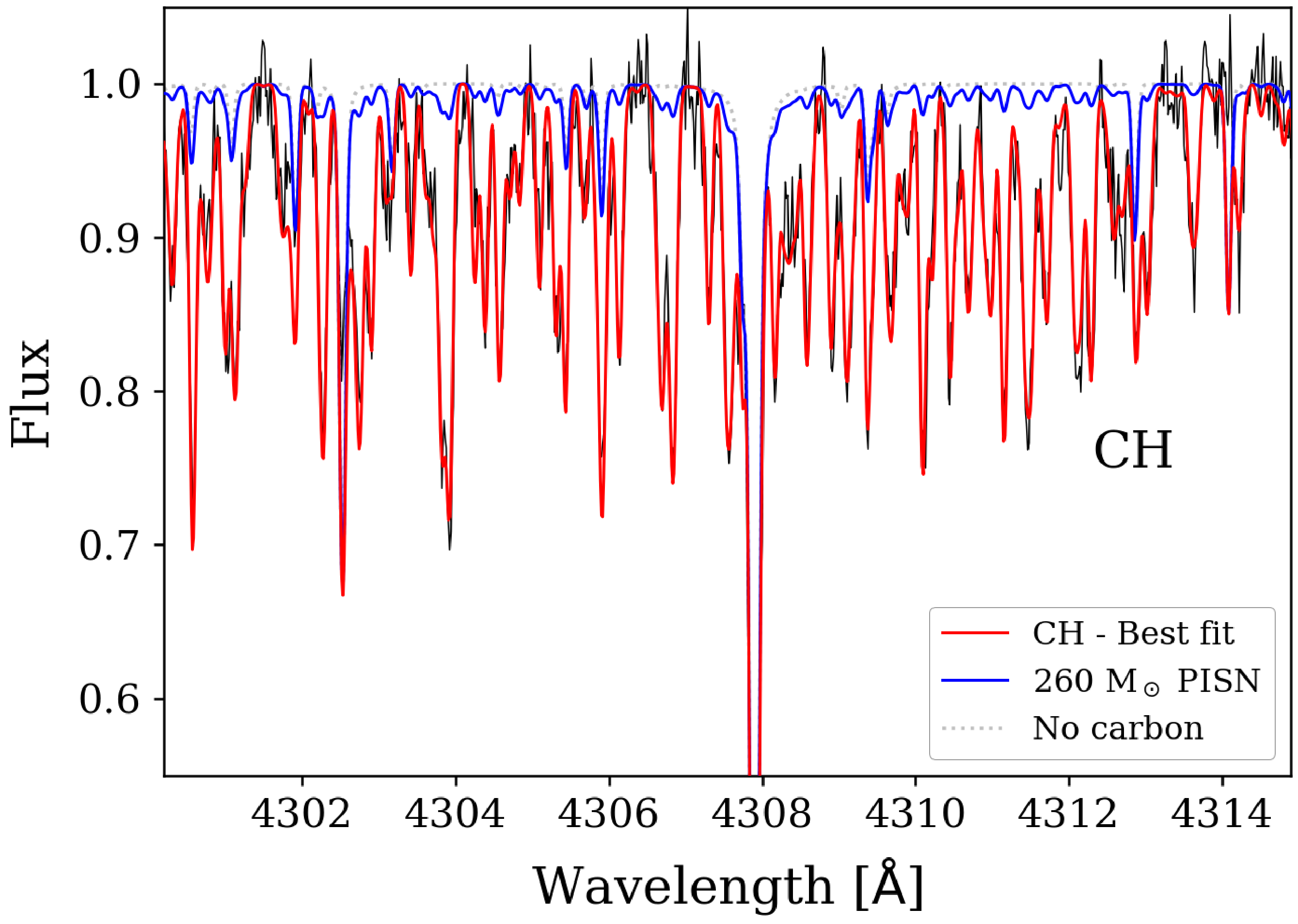}
\includegraphics[width=0.384\linewidth]{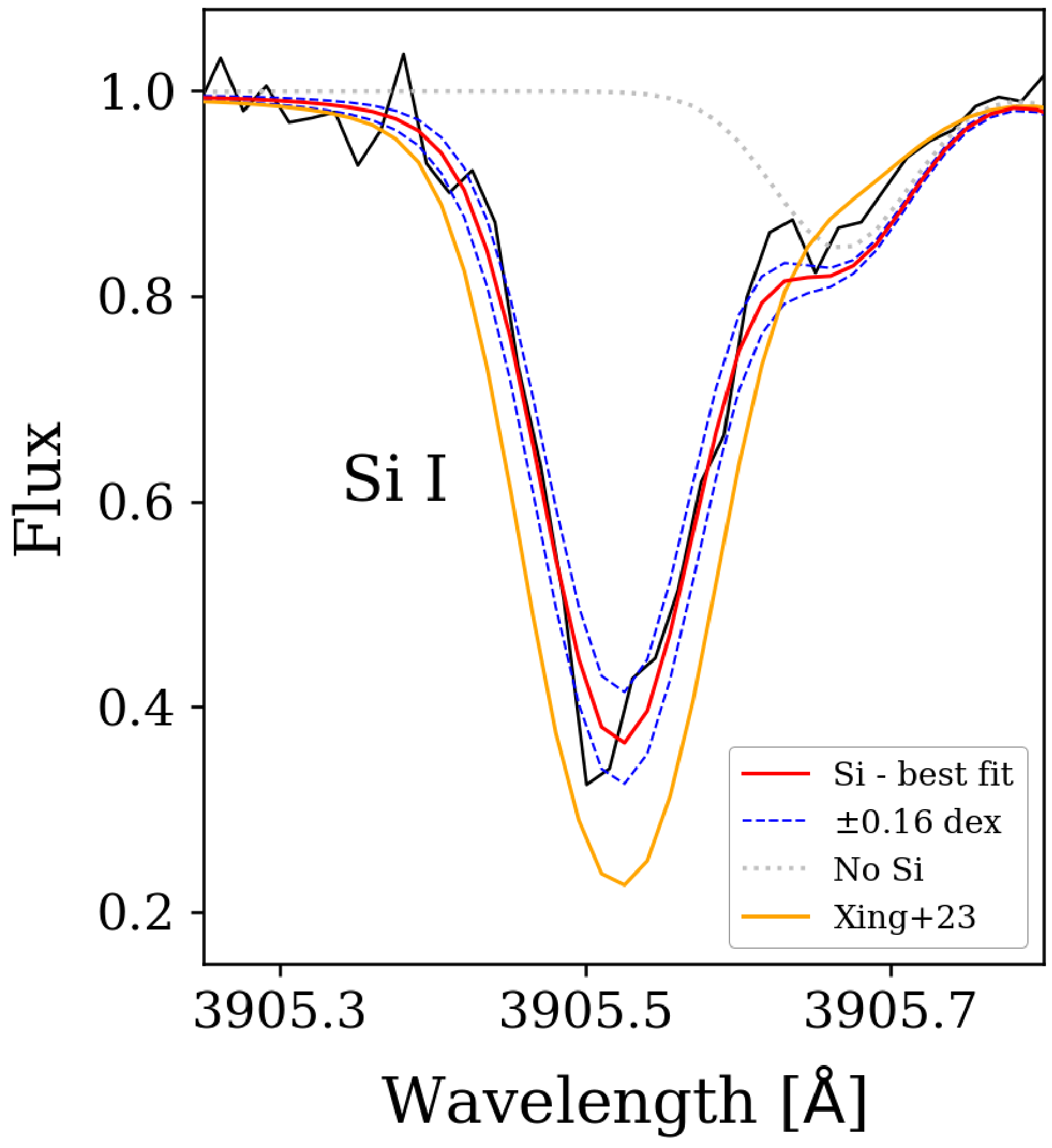}\\
\vspace{0.3cm}
\includegraphics[width=0.3\linewidth]{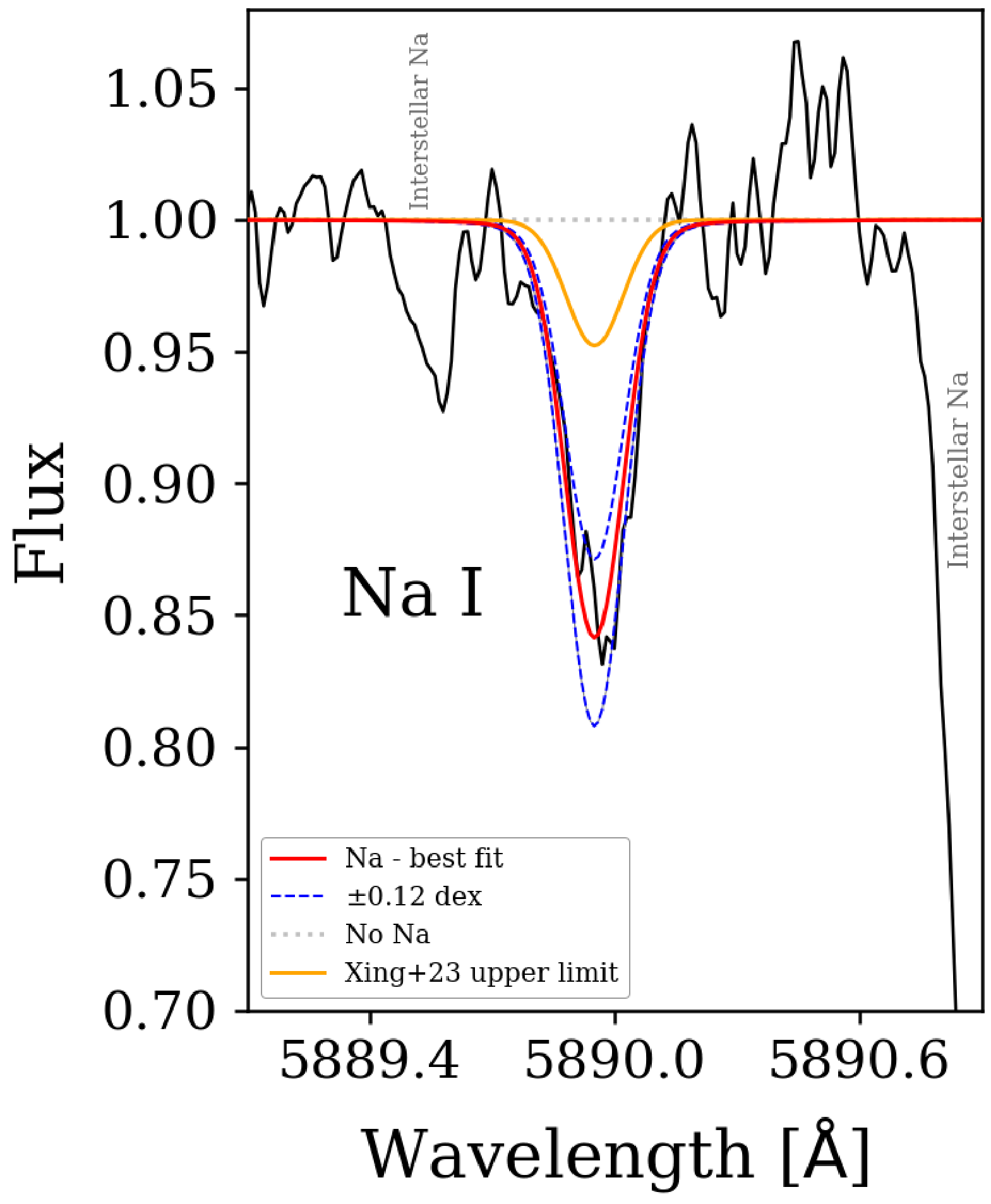}
\includegraphics[width=0.3\linewidth]{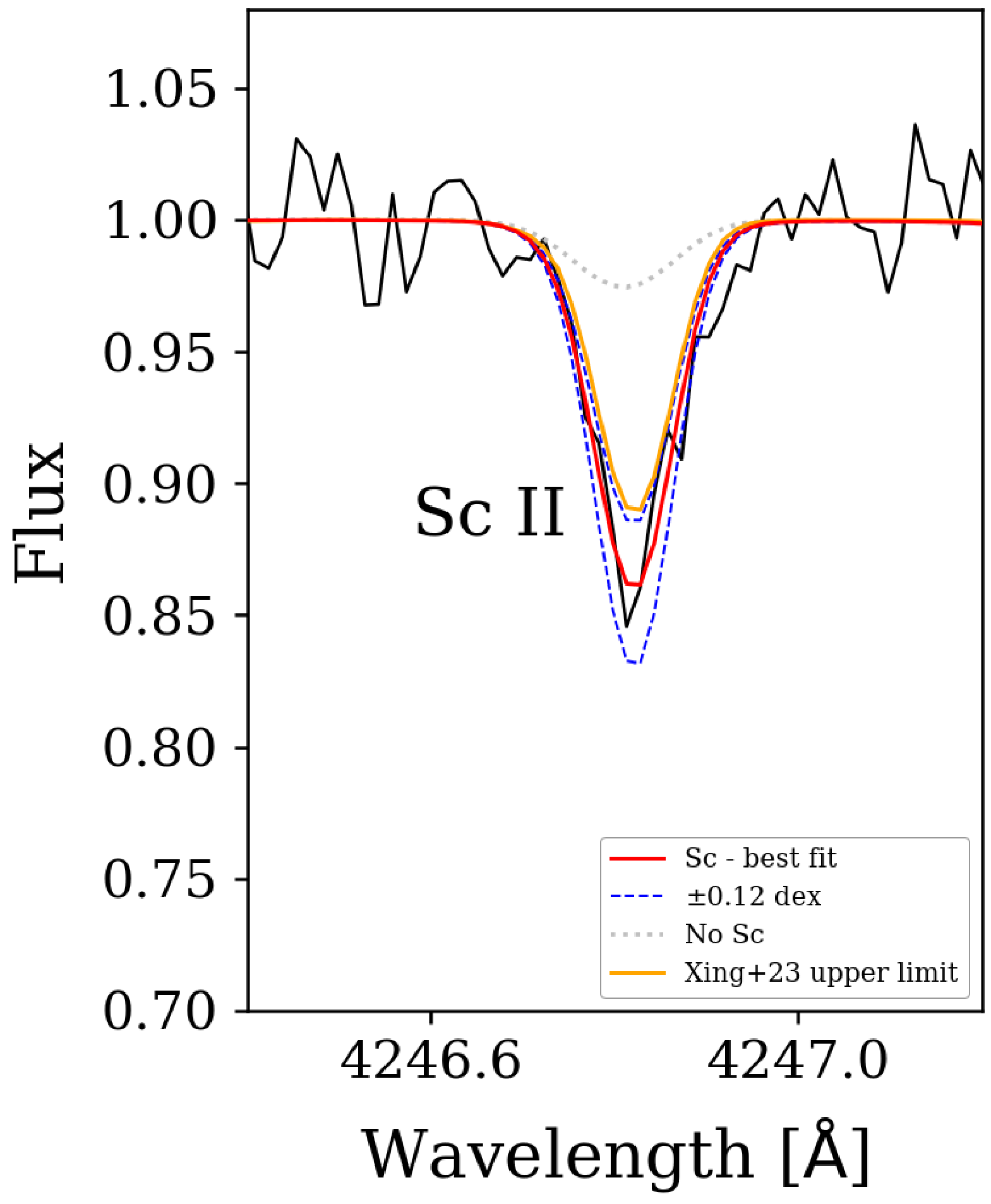}
\includegraphics[width=0.3\linewidth]{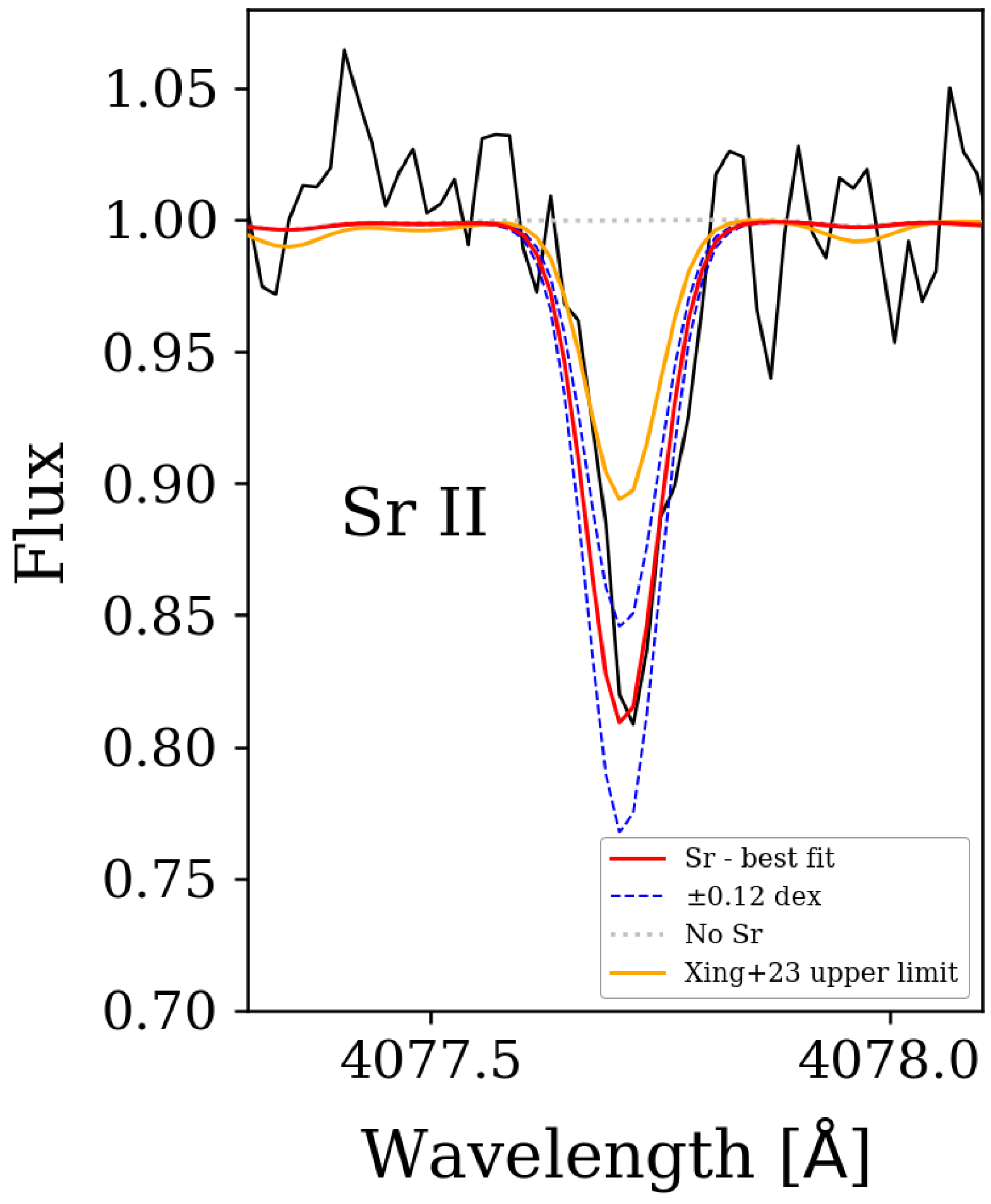}
\caption{Key spectral lines in the UVES spectrum (black) of J1010+2358. Synthetic spectra assume LTE and are shown in colors. Red are the best fits, and gray without the depicted element. Blue solid line (top left) shows the CH band if the star was a pure descendant of a 260\,M$_\odot$ PISN \citep{Heger02}, assuming the same 3D corrections for CH as adopted in Table~\ref{tab:abu}. Orange lines show synthetic spectra for the measurements (Si) or upper limits (Na, Sc, Sr) from \citet{Xing23}, by assuming their adopted stellar parameters (\teff=5860\,K, \logg=3.6, $v_{t}=1.5$\,km/s), but different Si lines were used in their analysis. All other depicted synthetic spectra (red, blue, gray) assume the stellar parameters of this work (\teff=5839\,K, $\log{g}$=4.72, $v_{t}=1.37$\,km/s).}
\label{fig:spectra}
\end{figure*}

\begin{figure}
\centering
\includegraphics[width=1\linewidth]{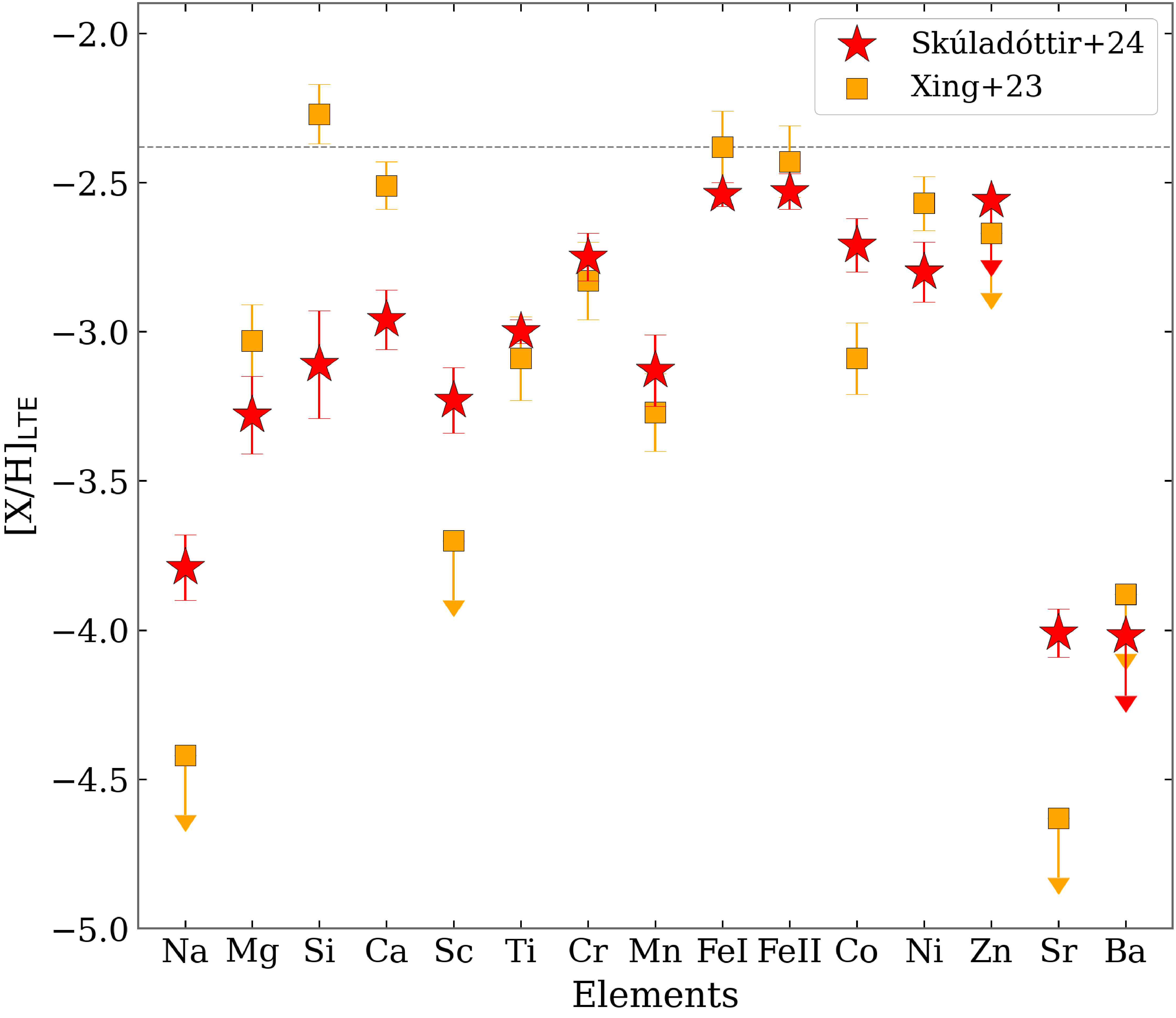}
\caption{Comparison between the chemical abundance pattern of J1010+2358 as measured by this work (red stars) and \citeauthor{Xing23} (\citeyear{Xing23}, orange squares). Dashed gray line marks the [\ion{Fe}{I}/H] of \citet{Xing23}; and for Ti, the abundances from \ion{Ti}{II} lines are shown. 
}
\label{fig:comp}
\end{figure}

\section{Chemical abundances}

The measured chemical abundances are listed in Table~\ref{tab:abu}, including errors, $\delta_{\rm[X/Fe]}$. The measured abundances were corrected for NLTE effects and/or 3D
effects by using grids available in the literature. Atomic data and results for individual lines are available in an online Table. 
The adopted solar scale is from \citet{Asplund21}, and literature results are adjusted accordingly. 
Here below we list some key aspects of the analysis (see more details in Sec.~\ref{sec:chemanalysis}). 

\subsection{Iron} \label{sec:iron}

The Fe abundance was measured from 179 lines of \ion{Fe}{i} and 6 lines of \ion{Fe}{ii}, at $376-496$\,nm. The \ion{Fe}{II} lines were corrected on a line-to-line bases using the 3D LTE grids from \citet{Amarsi19b}.  The 3D NLTE grid from \citet{Amarsi22} does not cover the parameters of this star; however, the results at the edge of this grid ($\log{g}_{\rm max}=4.5$) indicate that
the \ion{Fe}{II} lines used in this work show negligible departures from LTE.
This gave $\Delta\mathrm{[\ion{Fe}{II}/H]}_{\rm 3D\,NLTE}=+0.13$.
The corrections are similar for all the lines used here, and so the scatter between lines does not change after applying the corrections. 

For \ion{Fe}{I}, the 3D NLTE corrections of \citet{Amarsi22} for this star are higher, $\Delta$[\ion{Fe}{I}/{H}]$_{\rm 3D\,NLTE}=+0.24$. This gives rise to an ionization imbalance of $\rm [\ion{Fe}{I}/\ion{Fe}{II}]_{\rm 3D\,NLTE}=+0.10\pm0.07$.  However, the scatter between individual \ion{Fe}{I} lines is significantly increased, from $\sigma_{\text{LTE}}=0.10$ to $\sigma_{\text{NLTE}}=0.16$, driven by lines of intermediate excitation potential that feel the strongest NLTE effects.  As mentioned above, the adopted $\log{g}$ is outside the grid ($\log{g}_{\text{max}}=4.5$) and extrapolating the grid is likely to make the 3D NLTE corrections more uncertain. 

Therefore we deem the \ion{Fe}{ii} lines more reliable, and from here on out assume $\rm [Fe/H]=[\ion{Fe}{II}/H]_{\rm 3D\,NLTE}=-2.40\pm 0.06$. This is in good agreement with the LTE values reported in \citet{Xing23}, $\rm[\ion{Fe}{I}/H]_{\text{Xing}}=-2.38\pm0.12$, and $\rm [\ion{Fe}{II}/H]_{\text{Xing}}=-2.43\pm0.12$.

\subsection{Carbon and Aluminum}

The main goal of this project was to measure, with high precision and accuracy, the chemical abundances of C and Al which were not included in the study of \citet{Xing23}. These two elements have been shown to be fundamental to confirm or reject the interpretation that J1010+2358 is a pure descendant of a massive 260\,M$_{\odot}$ PISN \citep{Koutsouridou24,Jeena24}. 

The C was measured from both the CH band and a weak \ion{C}{I} line at 909.5\,nm. After applying 3D and/or NLTE corrections (\citealt{Amarsi19b,Norris19}, see Sec.~\ref{sec:cno}), both diagnostics agree with [C/Fe] being slightly subsolar (Table~\ref{tab:abu}), in stark contrast with the predicted yields of $\rm[C/Fe]_{\text{PISN, 260M$_\odot$}}<-1$ (see Fig.~\ref{fig:spectra}, top left).
The Al was measured from two neutral lines at 394 and 396\,nm, and the NLTE corrections were estimated from \citet{Nordlander17} to be $\Delta_{\rm NLTE}=+0.40$. The value of $\rm[Al/Fe]_{\rm NLTE}=-1.33\pm0.13$ is also significantly higher than that expected from a pure descendant of a massive PISN, $\rm[Al/Fe]_{\text{PISN, 260M$_\odot$}}<-2$. 

Only by looking at these two elements, C and Al we can thus exclude the previously proposed scenario where J1010+2358 is a pure descendant of 260\,M$_{\odot}$ PISN. 

\subsection{Comparison to previous results} \label{sec:comp}

Comparison to the abundance measurements of \citet{Xing23} is shown in Fig.~\ref{fig:comp}. We confirm the extremely low $\rm[\alpha/Fe]\lesssim-0.5$ for Mg and Ti which was reported by \citet{Xing23}. However, when it comes to several other elements we find significant discrepancies from their analysis. In particular, we do not find the high $\rm[Si/Fe]_{Xing,LTE}=+0.11\pm0.10$ they reported, but instead find much lower $\rm[Si/Fe]_{LTE}=-0.58\pm0.18$, similar to the other $\alpha$-elements, Mg and Ti. This corresponds to a difference of $\Delta\log\epsilon(\text{Si})_{\text{LTE}}=-0.84\pm0.21$ between the two analyses, see Figs.~\ref{fig:spectra} and~\ref{fig:comp}. By adopting the stellar parameters of \citet{Xing23}, we get $\Delta\log\epsilon(\text{Si})$ consistent with that presented in Table~\ref{tab:abu}. We note however, that the the \ion{Si}{I} line used here was outside the wavelength range of the Subaru spectrum and thus not included in the analysis of \citet{Xing23}. One of the lines they used, at 410.3\,nm (Xing, private communication), is not detected in our spectra, but provides an upper limit of $\Delta\log\epsilon(\text{Si})_{\text{LTE}}<4.95$, that is $\approx$0.3\,dex lower than the measured value of \citet{Xing23}. It is quite possible that unfortunately placed noise could have affected their measurement of this very weak line. In addition, the oscillator strength of the \ion{Si}{I} 410.3nm line is quite uncertain. NIST \citep{Ralchenko20} quote a value of $-3.34$\,dex based on theoretical calculations, with an uncertainty ranking of `E' ($>50\%$ uncertainty).  Recent experiments \citep{denHartog23} put the value 0.3 dex higher, $-3.03\pm0.08$\,dex, corresponding to around a 0.31 dex reduction in inferred silicon abundance. In addition, we find a lower Ca compared to \citet{Xing23}, $\Delta\log\epsilon(\text{Ca})_{\text{LTE}}=-0.45\pm0.13$. The lower Si and Ca we measure, both reduce the strong odd-even effect found by \citet{Xing23}, which is predicted to be characteristic for PISN yields \citep[e.g.][]{Heger02}.

In three cases where \citet{Xing23} only reported upper limits (Na, Sc and Sr), we were able to unambiguously detect the lines with our higher-resolution UVES spectra. Furthermore, our measurements for these three elements are $\gtrsim 0.5$\,dex higher than the reported upper limits, see Fig.~\ref{fig:spectra} and~\ref{fig:comp}. In the case of Na and Sc this again greatly reduces the odd-even pattern of J1010+2358. 
The described discrepancies cannot be explained only with the differences in adopted stellar parameters. Fig.~\ref{fig:spectra} (orange) shows synthetic spectra for the measured abundances and upper limits of \citet{Xing23}, by assuming their adopted stellar abundances, and the UVES spectrum is not consistent with their results.



\section{Origin of J1010+2358}

\subsection{Fitting procedure}

Given the unique abundance pattern of J1010+2358, we assume it has been enriched by either one or two progenitors, that can be of any type, i.e., PISNe, Pop~III SNe (ranging in mass, energy and internal mixing), Pop~II ccSNe (ranging in mass and metallicity) or type~Ia SNe (SNIa, ranging in metallicity and initial central density). We adopt the yields of \citet{Heger02,Heger2010} for Pop~III stars and consider both the \citet{Woosley1995} and \citet{Limongi2018} yields for Pop~II ccSNe; and the \citet{Iwamoto1999} and \citet{Leung2018} yields for SNIa. The fitting follows the approach of \citet{Vanni2024}, details in Sec.~\ref{sec:fittingdetails}, with the difference that here models are excluded that violate upper or lower limits of measured chemical abundances.


For the fit, we use the measured NLTE abundances (Table~\ref{tab:abu}), and adopt the \ion{C}{I} abundance for C (Sec.~\ref{sec:cno}). Since the NLTE effects for Al and Ni are uncertain~(Sec.~\ref{sec:fepeak}), we adopt larger errors, $\delta_{\text{[Al/Fe]}}=0.3$ and $\delta_{\text{[Ni/Fe]}}=0.25$. Finally, the NLTE corrections for Co \citep{Bergemann10Co} are exceptionally large, $\rm\Delta NLTE=+0.56$, and likely to be uncertain.
For a conservative approach, we assume $\rm\Delta NLTE(Co)>0$, and adopt the LTE value, including the error, as a lower limit, $\rm[Co/Fe]_{NLTE}>[Co/Fe]_{LTE}-\delta_{\text{[Co/Fe]}}$. Otherwise the measured NLTE abundances and errors are as listed in Table~\ref{tab:abu}.

\begin{figure}
    \centering
    \includegraphics[width=1.0\linewidth]{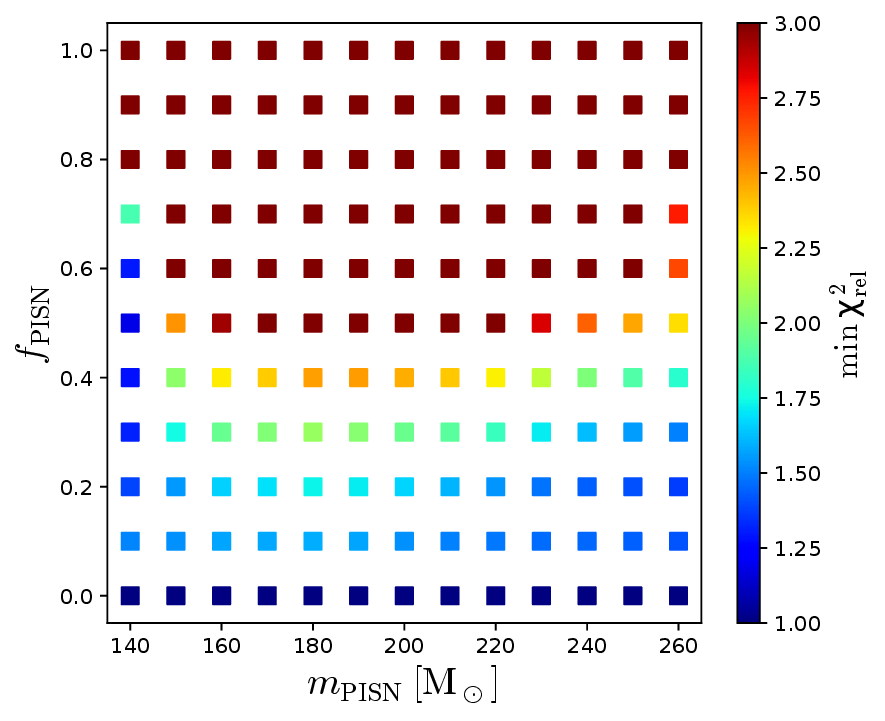}
    \caption{Minimum $\chi^2_{\rm rel}$ over all fits containing two progenitors: i)~a PISN of a given mass $m_{\rm PISN}$ and fraction of enrichment, $f_{\rm PISN}$, ii) another star providing the remaining metals.}
    \label{fig:f_pisn}
\end{figure}

\begin{figure*}
    \centering
    \includegraphics[width=1.0\linewidth]{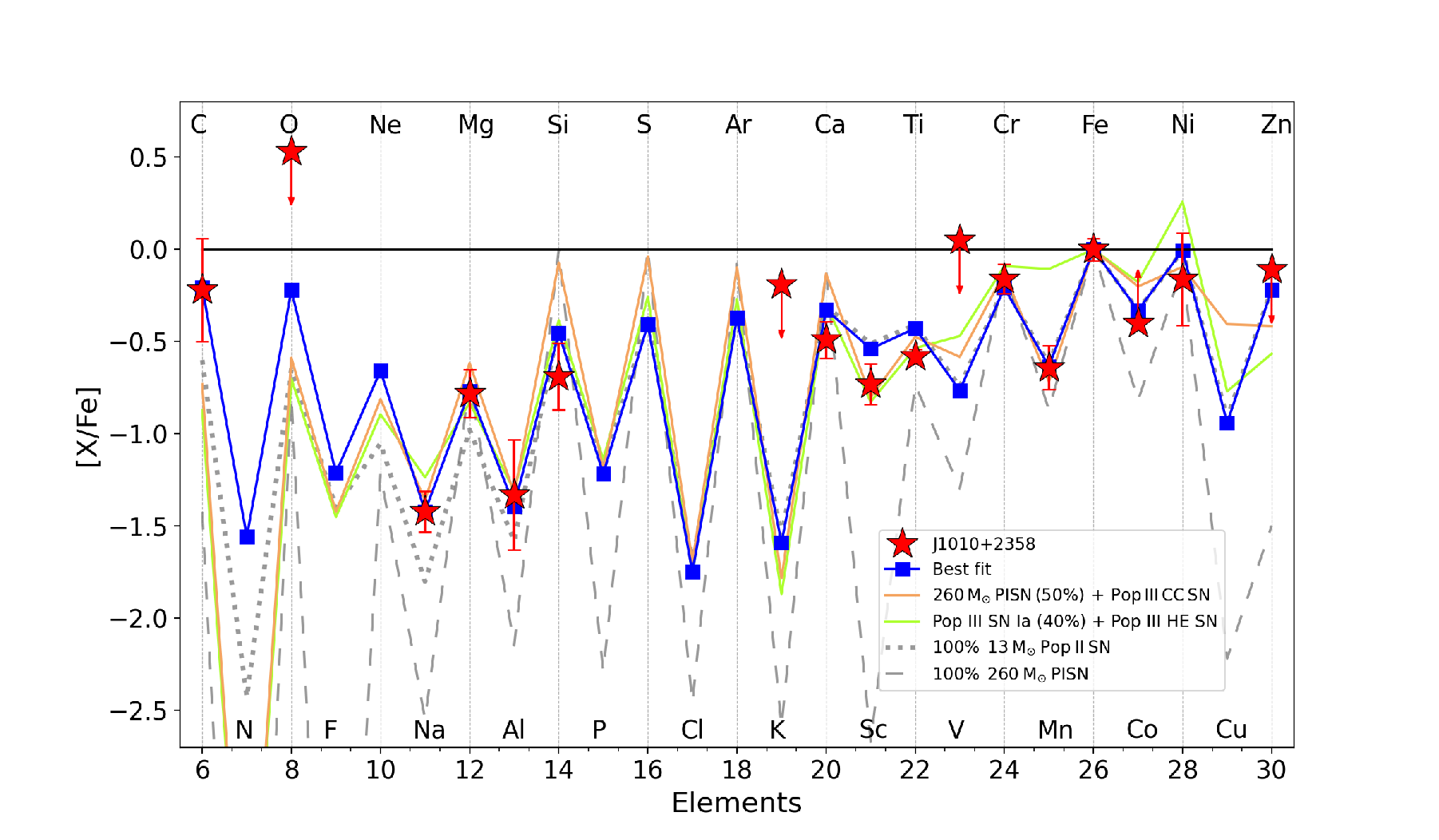}
    \caption{Abundance pattern of J1010+2358 (red) compared with the best fit (blue), where 50\% of the metals originate from a 13\:${\rm M_\odot}$ Pop~II ccSN, and the rest comes from a Pop~III ccSN ($m_\star$=39 ${\rm M_\odot}$, $\rm E_{SN}=0.9 \times 10^{51}\:$erg). Additionally shown are environments enriched by: (i)~a $260 \rm M_{\odot}$ PISN (50\%) and a Pop~III core-collapse SN (orange); (ii)~a Pop~III SN Ia (40\%) and a high-energy Pop~III SN (green); (iii)~the $13 \rm M_{\odot}$ Pop~II SN (100\%) which is present in all our best fits ($\chi^2_{\rm rel}<2$); and (iv)~a $260 \, $M$_{\odot}$ PISN (100\%), as the originally proposed origin of J1010+2358 by \citet{Xing23} (gray dashed).}
    \label{fig:chi_squares}
\end{figure*}

\subsection{Best fits}
\label{sec:bestfits}

First, we examine the possibility of J1010+2358 having inherited a fraction of its metals, $f_{\rm PISN}$, from a PISN of mass $m_{\rm PISN}$; while the rest of metals comes from a second star of any type. Following the approach of \citet{Koutsouridou24}, Fig.~\ref{fig:f_pisn} shows the best fits (the minimum $\chi^2_{\rm rel} \equiv \chi^2/\chi^2_{\rm best}$) for a given $m_{\rm PISN}$ and $f_{\rm PISN}$, over all possible secondary metal-sources.

Notably, even a 10$\%$ contribution from a PISN reduces the fit quality to $\chi^2_{\rm rel}>1.4$. Nonetheless, contributions of $\leq 40\%$ from a $150-260\,{\rm M_\odot}$ PISN or a $\leq 70\%$ from a 140$\,{\rm M_\odot}$ PISN cannot be conclusively ruled out since $\chi^2_{\rm rel}<2$. Crucially, in {\it all} cases of $\rm min(\chi^2_{\rm rel})<2$ in Fig.~\ref{fig:f_pisn}, the second SN contributing the remaining metals, is a  13\,${\rm M_\odot}$ Pop~II ccSN \citep{Woosley1995}. The preferred metallicity of this low-mass Pop~II SN varies in the best fits, $Z = [10^{-4}, 10^{-1}]\:Z_\odot$, depending on the properties of the other SN that contributes the remainder of the metals in J1010+2358.

This same star is also included in the overall best fit, $\chi^2_{\rm best}=13.2$ (blue line in Fig.~\ref{fig:chi_squares}), which is obtained when half of the metal content of J1010+2358 originates from this 13\:${\rm M_\odot}$ Pop~II ccSN ($Z=10^{-2}\:Z_\odot$), using the yields of \citet{Woosley1995}, and the remaining half comes from a Pop~III ccSN ($m_\star$=39 ${\rm M_\odot}$, $\rm E_{SN}=0.9 \times 10^{51}\:$erg). In addition, this star is consistently present in all other solutions with $\chi^2_{\rm rel}<2$. 
Contrarily, we find {\it no solutions} with $\chi^2_{\rm rel}<2$ that involve only Pop~III stars, or a Pop~II ccSN of higher mass ($m_\star\geq 15\,$M$_\odot$) when using the yields of \citet{Woosley1995}, or any mass when using the yields of \citet{Limongi2018}. Clearly, the contribution from this 13\,${\rm M_\odot}$ Pop~II ccSN is pivotal. As shown in Fig.~\ref{fig:chi_squares} (gray dotted line), the predicted abundances closely resemble the observations, albeit with slightly lower abundances of the lighter elements ($Z\leq12$). Thus there are many good fits ($\chi^2_{\rm rel}<1.4$) with this 13\,M$_\odot$ Pop~II ccSN and a Pop~III SN, which commonly have high yields of lighter elements (e.g.~\citealt{Vanni23}). 

We note that this peculiar abundance pattern is not seen in the low-mass Pop~II ccSNe of \citet{Limongi2018}, nor in the low-mass Pop~III ccSNe of \citet{Heger02}. However, \citet{Jeena24} proposed an alternative explanation for J1010+2358, namely a low-mass Pop~III ccSN. Their yield predictions (shown in their Fig.~3) would not provide good fits to our measured abundances of J1010+2358 as their Si is significantly higher, and the Na and Sc are predicted to be $\gtrsim0.6$\,dex lower.

To explore alternative solutions, we excluded the \citet{Woosley1995} 13\:${\rm M_\odot}$ Pop~II ccSN from our fitting routines. In this scenario, two possible optimal fits emerge ($\chi^2_{\rm rel}\approx2.3$); the first involving a $50\%$ contribution from a Pop~III ccSN paired with a 260$\:{\rm M_\odot}$ PISN (orange line in \ref{fig:chi_squares}); and the second featuring a 60$\%$ high energy Pop~III SN paired with a zero-metallicity SNIa (green line; W70 model by \citealt{Iwamoto1999}). The Pop~III SNIa yields have a stronger odd-even effect relative to their Pop~II counterparts, resulting in a better fit with J1010+2358. We note that the main discrepancy of this scenario is in the Mn abundance (Fig.~\ref{fig:chi_squares}), which is sensitive to the mass of the white dwarf \citep[e.g.][]{Nissen24}. So it is likely that a comparable model of a Pop~III SNIa, but with a sub-Chandrasekhar mass white dwarf instead of near-Chandrasekhar, would provide a better fit.

\section{Discussion and Conclusions}

With the aim to verify that J1010+2358 is a pure descendant of a 260\,M$_{\odot}$ PISN, as proposed by \citet{Xing23}, we obtained a high-resolution, high-quality VLT/UVES spectrum of the star. Our main goal was to measure C and Al, as these elements are fundamental to quantify the fraction of metals, $f_{\rm PISN}$, coming from a massive PISN  \citep{Koutsouridou24}. The measured $\rm[C/Fe]_{\rm 3D\,NLTE}$ is more than $1$\,dex higher than the value predicted for a 260\,M$_{\odot}$ PISN, and $\rm[Al/Fe]_{NLTE}$ is also $0.8$\,dex higher. The C and Al abundances alone can therefore robustly exclude that J1010+2358 obtained all of its metals from a 260\,M$_{\odot}$ PISN. 

For other elements, several discrepancies were found in the abundance analysis of this work relative to that of \citet{Xing23}, see Fig.~\ref{fig:spectra}-\ref{fig:comp} and Sec.~\ref{sec:comp}. In particular, for three elements, Na, Sc, and Sr, our clear detection of the relevant absorption lines resulted in abundances that were $\gtrsim{0.5}$\,dex higher than the upper limits reported by \citet{Xing23}. In addition, our $\rm \log \epsilon(Si)_{LTE}$ is 
$\approx0.8$\,dex lower than that derived by \citet{Xing23}. The odd-even effect in J1010+2358 is therefore much less pronounced compared to \citet{Xing23}.

Explaining the discrepancies to \citet{Xing23} is nontrivial.  We do not have reason to suspect significant problems with the observed spectrum: key regions of the Subaru spectrum are shown in \citet{Thibodeaux24}, where the CH band is clearly visible, and the Na line is detectable. \citet{Thibodeaux24} also report an analysis of  J1010+2358 using high-resolution Keck/HIRES spectrum, getting results very consistent with our abundance pattern (Table~\ref{tab:abu}). They find C and Al to be higher than what is expected for a pure descendant of a PISN and also detect the lines of Na, Sc, and Sr at significantly higher values than reported in \citet{Xing23}, but Si is not included in their analysis. We speculate that unfortunately placed noise might influence the high Si measured in \citet{Xing23}, and differences in linelists and atomic data could explain some of the discrepancies (Sec.~\ref{sec:comp}). However, for many species it is unclear exactly which lines and atomic data were used by \citet{Xing23}, which prevents us from doing a more detailed comparison.\\
\\

To understand the origin of J1010+2358, we compare its unique abundance pattern with the predicted yields of one or two stars of any kind (PISNe, Pop~III/Pop~II SNe, or type~Ia SNe). The best fit, $\chi^2_{\rm best}$, includes a combination of a 13\,M$_{\odot}$ Pop~II ccSN and a Pop~III ccSN of 39\,M$_{\odot}$. We are able to find many fits of comparable goodness, but \textit{all} fits with $\chi^2_{\rm rel}=\chi^2/\chi^2_{\rm best}\leq2$ include the 13\,M$_{\odot}$ Pop~II ccSN from \citet{Woosley1995}. This low-mass ccSN has very unique chemical yields that closely resemble the abundance pattern of J1010+2358 (Fig.~\ref{fig:chi_squares}).
Two alternative solutions ($\chi^2_{\rm rel}\approx2.3$) were found by excluding the 13\,M$_\odot$ Pop~II ccSN from the fit: i)~50\% of the metals coming from a 260\,M$_\odot$ PISN; ii)~40\% of the metals coming from a zero-metallicity SNIa. 

All the proposed scenarios have some drawbacks. It is unclear how reliable the yields are for the 13\,M$_\odot$ Pop~II ccSN from \citet{Woosley1995}, given that other works are not in agreement (Sec.~\ref{sec:bestfits}). In addition, it seems quite unlikely that a single Pop~II managed to enrich an environment otherwise only polluted by Pop~III stars. For the other proposed solutions the fits are significantly worse ($\chi^2_{\rm rel}\approx2.3$). In the case of a 40\% contribution from Pop~III SNIa, it is expected that Pop~III AGB stars would already have enriched the environment, but the low $\rm[Ba/H]\lesssim-4$ may rule this out. 
In any case, J1010+2358 is clearly a rare object with a unique abundance pattern, and so unlikely scenarios cannot be excluded.

Ultimately, a dominant PISN contribution of $\gtrsim70\%$ to the metals of J1010+2358, can be excluded. Lower, non-zero PISN contribution cannot be ruled out, but is not necessarily favoured (Fig.~\ref{fig:f_pisn}). This means, unfortunately, that our results do not support the previous claim that all the metals of J1010+2358 are inherited from a PISN. Even if this star possibly contains some PISN contribution, it is likely too low to be useful for constraining the Pop~III IMF \citep{Koutsouridou24}. Our results emphasize the need for careful chemical abundance analyses, fully taking into account possible 3D and/or NLTE effects to robustly identify a true PISN descendant. Armed with the lessons learned from J1010+2358, both from an observational and theoretical point of view, we are now equipped to handle future candidates for PISN enrichement which will likely be revealed by the upcoming large spectroscopic surveys.\\
\\

{\footnotesize \noindent This project has received funding from the European Research Council (ERC) under the European Union’s Horizon 2020 research and innovation programme (grant agreement No. 101117455). I.V., I.K, R.L, and S.S. acknowledge funding from the ERC grant No. 804240.  A.M.A acknowledges support from the Swedish Research Council (VR 2020-03940).  D.A. acknowledges financial support from the Spanish Ministry of Science and Innovation (MICINN) under the 2021 Ram\'on y Cajal program MICINN RYC2021‐032609. We thank the anonymous referee for a very insightful input that helped improve this work.}\\
\\
{\footnotesize
{\footnotesize \noindent{\it Facilities:} Based on observations made with ESO VLT at the La Silla Paranal observatory under program ID 112.26WJ.001. This work has made use of data from the European Space Agency (ESA) mission Gaia (https://www.cosmos.esa.int/gaia), processed by the Gaia Data Processing and Analysis Consortium (DPAC, https://www.cosmos.esa.int/web/gaia/dpac/ consortium). Funding for the DPAC has been provided by national institutions, in particular the institutions participating in the Gaia Multilateral Agreement.}}

\bibliography{2pisnornot2pisn}{}
\bibliographystyle{aasjournal}

\appendix

\section{Isochrone fitting}

\begin{figure}
\centering
\includegraphics[width=0.8\linewidth]{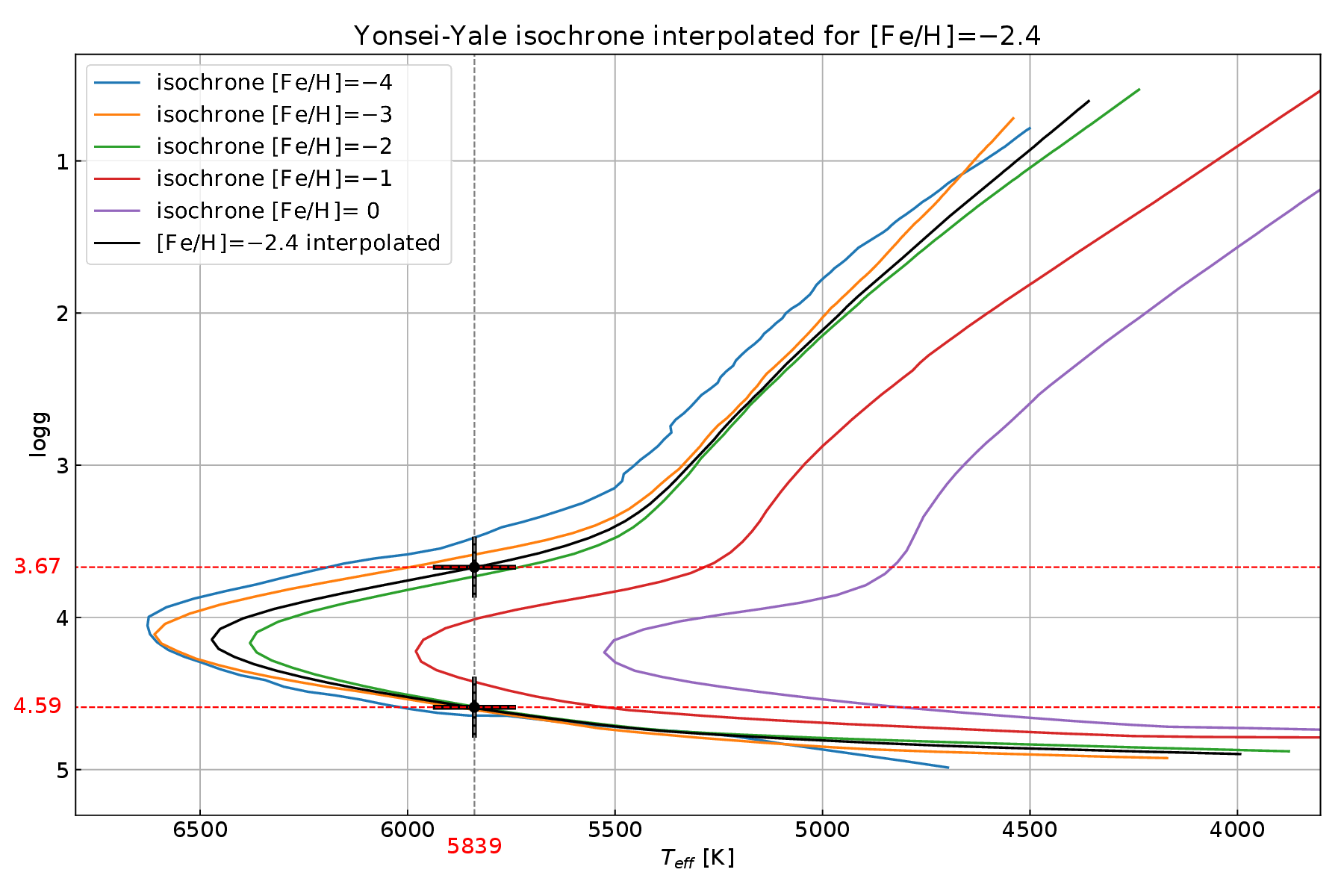}
\caption{Yonsei-Yale isochrones for stars with $M_\star=0.8$\,M$_{\odot}$ and age 13~Gyr. Black line shows the interpolated isochrone for a star with $\rm[Fe/H]=-2.4$, and black points for our adopted temperature of $\teff=5839$\,K. Enhancement of $\rm [\alpha/Fe]=+0.3$ is assumed, except in the case of [Fe/H]=0 where $\rm[\alpha/Fe]=0$.}
\label{fig:iso}
\end{figure}

Figure \ref{fig:iso} shows the \logg\ for J1010+2358 according to the Yonsei-Yale isochrones \citep{Demarque04}, by adopting $\teff=5839$\,K. For a subgiant this would correspond to \logg=3.67, in good agreement with the value assumed by \citet{Xing23}, $\log{g}(\text{Xing})=3.6\pm0.2$. On the other hand when the star is assumed to be a dwarf, the isochrone fitting gives \logg=4.59, consistent with the photometric \logg\ adopted by this work, $\log{g}=4.72\pm0.13$, which is based on the distance to the star as estimated by Gaia DR3 photometry \citep{Bailer-Jones21}. No isochrones are available with $\rm[\alpha/Fe]<0$ and $\rm[Fe/H]<0$ as is measured in J1010+2358. However, since the isochrone values are only used for confirmation of the adopted photometric values, this should not affect this work.

\section{Details of chemical abundance analysis} \label{sec:chemanalysis}

A comparison between our LTE results and those of \citet{Xing23} are shown in Fig.~\ref{fig:comp}. Most notable differences are the $\alpha$-elements Si and Ca, which have significantly lower abundances in the analysis of this work; and the three upper limits reported for Na, Sc, Sr by \citet{Xing23}, while here we detect the absorption lines of these elements, and derive $\gtrsim0.5$\,dex higher abundances (Fig.~\ref{fig:spectra}). Below are listed the details of the LTE abundance analysis and the applied 3D and/or NLTE corrections.

\subsection{Lithium}

The Li was measured from the \ion{Li}{i} doublet at 670.7\,nm, and NLTE corrections were applied from \cite{Lind09}. The value, $\rm \log
\epsilon(Li)_{\text{NLTE}}=1.90\pm0.10$ falls slightly below the Spite plateau at $\rm\log\epsilon(Li)=2.2$ \citep{Spite82, Rebolo88}, indicating a possible minor depletion but in line with what is observed for metal-poor dwarfs spanning metallicities from $\rm [Fe/H]=-2$ to $-6$ \citep[see e.g.,][]{Aguado19}.

\subsection{Carbon, nitrogen and oxygen} \label{sec:cno}

The C was measured both from CH and the \ion{C}{I} line at 909.5\,nm. \citet{Xing23} did not report a measurement or an upper limit of CH, but in our UVES spectra the detection of the CH G-band at $\sim$430\,nm is unambiguous (Fig.~\ref{fig:spectra}). The CH abundance, $\rm [CH/Fe]_{\rm 3D}=-0.06\pm0.14$, was based on an average from wavelength windows in the range 423-439\,nm, and the measurements agreed very well over the region, $\sigma_{\text{CH}}=0.04$. In the absence of any measurable O lines we adopted $\rm [O/Fe]=0$ when synthesising the CH-region. Given the overall low $\rm[\alpha/Fe]<0$ in this star, it is likely that the O is also subsolar. But assuming $\rm[O/Fe]=-0.5$ instead only results in $\rm \Delta[CH/Fe]=-0.03$, so the expected change based on the true O abundance is well within the error bars. From the work of \citet{Norris19} we estimate a 3D correction of $\rm \Delta [CH/H]_{3D}=-0.35$, but we note that this correction is quite uncertain, as seen by their Fig.~2. The \ion{C}{I} line at 909.5\,nm was in between telluric lines at the time of observations, and 3D NLTE corrections were applied according to \citep{Amarsi19a,Amarsi19b}, giving $\rm [\ion{C}{I}/Fe]_{\rm 3D}=-0.22\pm0.28$. We get reasonable agreement between the two diagnostics, $\rm [\ion{C}{I}/CH]_{\text{3D NLTE/3D}}=-0.16 \pm0.31$. The measured [C/Fe] is therefore consistent with solar, which is typical for C-normal metal-poor halo stars, $\rm[\ion{C}{I}/Fe]^{\text{halo}}_{\text{3D NLTE}}\approx0$ at $\rm[Fe/H]<-2$ \citep{Amarsi19b}.

No O lines were detected in the UVES spectrum, and an upper limit was derived based on the 777.2\,nm line. Corrections for 3D NLTE effects were based on \citet{Amarsi19b}. The upper limit of $\rm[O/Fe]_{\text{3D NLTE}}<+0.53$, is an indicator that this star is likely lower in O compared to the general halo population, where at $\rm[Fe/H]<-2$, the average is $\rm[O/Fe]^{\text{halo}}_{\text{3D NLTE}}=0.62$ \citep{Amarsi19b}. The CN bandhead at 388\,nm is not detected, giving an upper limit of $\rm [N/Fe]<+1.25$.

\subsection{Light odd elements: Na, Al, K, Sc}

The two \ion{Na}{I} lines around 589\,nm were detected, see Fig.~\ref{fig:spectra}, and gave very consistent results, agreeing within 0.02\,dex. The NLTE effects are expected to be within the measurement errors, and were adopted from \citet{Lind11}. The measured value of $\rm[Na/Fe]=-1.42\pm0.11$ is not in agreement with \citet{Xing23} who were not able to detect these \ion{Na}{I} lines, but reported an upper limit of $\rm[Na/Fe]_{\text{Xing}}<-2.04$. Like \citet{Xing23} we also detect strong interstellar Na absorption red of the stellar lines, as well as several weaker interstellar absorption lines in the vicinity.

The Al abundance was measured from two lines at 394 and 396\,nm, which agreed within 0.07\,dex. The NLTE corrections were estimated from \citet{Nordlander17} to be $\rm\Delta[Al/H]_{\text{NLTE}}=+0.4$. The \ion{K}{I} line at 770\,nm was blended with a telluric line slightly shifted to the blue relative to the line. The unblended red wing of the line was therefore used to estimate an upper limit, and NLTE effects were adopted from \citet{Reggiani19}. Neither Al nor K were measured by \citet{Xing23}.

The \ion{Sc}{II} line at 427\,nm gave $\rm[Sc/Fe]_{NLTE}=-0.73\pm0.11$. For the NLTE effects on \ion{Sc}{II} we rely on the work of \citet{Mashonkina22}, where stars with stellar atmospheric parameters similar to J1010+2358 have corrections on the order of $+0.1$\,dex. The detection of Sc in the UVES spectrum does not agree with the upper limit provided by \citet{Xing23}, $\rm[Sc/Fe]_{\text{Xing}}<-1.32$, see Fig.~\ref{fig:spectra}.

\subsection{The $\alpha$-elements: Mg, Si, Ca, Ti}

The Mg abundance was measured from the \ion{Mg}{I} triplet at 383\,nm, and all three lines agreed within 0.02\,dex. The 3D NLTE corrections were calculated for the \ion{Mg}{I} triplet as a whole, using the methods described in \citet{Nissen24} and Matsuno et al. in prep. Although we used different lines, our measured Mg abundance is in good agreement with that of \citet{Xing23}.

The \ion{Si}{i} line around 391\,nm gave the result $\rm[Si/Fe]_{\text{NLTE}}=-0.69\pm0.18$, adopting very minor corrections of $\Delta_{\rm NLTE}=+0.02$ from \citet{Amarsi17}\footnote{\url{https://www.astro.uu.se/~amarsi/}}.
This is in stark contrast to the value $\rm[Si/Fe]_{\text{Xing}}=+0.11\pm 0.12$ obtained by \citet{Xing23}, see Fig.~\ref{fig:spectra}. We note that \citet{Xing23} did not have access to the \ion{Si}{I} line used here.

Four neutral lines were used to measure Ca, at 423, 430, 443, and 445\,nm, which agreed well, $\sigma_{\text{Ca}}=0.03$. The MPIA database was used to estimate the NLTE effects \citep{Mashonkina07}. Our measured abundance of $\rm[Ca/Fe]_{\text{NLTE}}=-0.49\pm0.10$ is somewhat lower than that of \citet{Xing23}, $\rm[Ca/Fe]_{\text{Xing}}=-0.13\pm0.08$, possibly in part due to different line selection. However, \citet{Xing23} does not list what Ca lines were used, hindering a more detailed comparison.

The Ti was based on 11 lines of the ionized species which are less sensitive to NLTE effects, compared to the neutral species. Based on \citet{Mallinson24}, we estimate a correction of $\Delta$[\ion{Ti}{II}/H]$_{\text{NLTE}}=+0.02$. Our sub-solar Ti abundance is in good agreement with that of \citet{Xing23}.

\subsection{Iron-peak elements: V, Cr, Mn, Co, Ni, Zn} \label{sec:fepeak}

The upper limit of Vanadium was based on two \ion{V}{I} lines around 438\,nm. No detailed study of the NLTE effects on vanadium lines exist, but \cite{Ou20} find a difference of [\ion{V}{II}/\ion{V}{I}$]=+0.25$, and argue that this is most likely due to 
NLTE effects on the lines of \ion{V}{I}. We therefore increase our upper limit based on \ion{V}{I} lines accordingly.

The NLTE corrections for \ion{Cr}{I} are adopted from \citep{Bergemann10Cr}. Four \ion{Mn}{I} lines gave the value $\rm[Mn/Fe]_{NLTE}=-0.64\pm0.12$, with NLTE corrections from \citet{Bergemann19}. Both Cr and Mn we measure here agree well with \citet{Xing23}, $|\Delta\rm [Mn,Cr/H]_{\text{LTE}}|<0.20$.

The NLTE corrections for Co are quite large, $\rm\Delta[Co/H]_{\text{NLTE}}=+0.56$ \citep{Bergemann10Co}.
However, the LTE abundance of this work (TW) is somewhat higher than that of \cite{Xing23} with $\rm\Delta[Co/H]^{\text{TW-Xing}}_{\text{LTE}}=+0.38\pm0.15$. The Ni abundance was based on 5 \ion{Ni}{I} lines. There are not any NLTE corrections for Ni available in the literature, and therefore we adopt the same correction as for \ion{Fe}{I}, as both of these species are expected to suffer from over-ionisation. The non-detection of the \ion{Zn}{I} line at 481\,nm, yielded an upper limit of Zn, to which we estimated NLTE corrections of $\rm \Delta[Zn/H]_{\rm NLTE}=+0.05$ \citep{Takeda05}.

\subsection{Neutron-capture elements: Sr, Ba}

Contrary to \citep{Xing23}, we were able to detect two \ion{Sr}{II} lines at 408\,nm and 422\,nm, which gave $\rm [Sr/Fe]_{\text{LTE}}=-1.48\pm0.08$, with $\sigma_{\text{Sr}}=0.06$. 
Our measured Sr abundance is in disagreement with the upper limit provided by \citet{Xing23}, $\rm[Sr/Fe]_{\text{Xing}}<-2.25$, see Figs.~\ref{fig:spectra} and~\ref{fig:comp}. No Ba line was detected in our spectrum, and the upper limit was determined from the 455\,nm line. Our upper limit for Ba is similar to that that provided by \citep{Xing23}, see Fig.~\ref{fig:comp}. The NLTE corrections for Sr and Ba were estimated based on the work of \citet{Mashonkina23} to be $\rm \Delta[Sr/H]_{\rm NLTE}=+0.15$ and $\rm \Delta[Ba/H]_{\rm NLTE}=+0.11$.

\subsection{Error estimation}\label{sec:error}

For all measured elemental abundances,  we estimate the error as the quadratic sum of the random error of the mean, $\sigma / \sqrt{N_{l}-1}$; and the error arising from the stellar parameters. In all cases where we had 2-3 lines, they agreed very well with scatter $\sigma\lesssim0.03$, which is not representative of the true error. In the cases of $N_l\leq3$ we therefore adopted the error of an individual line as $\sigma$. Not included in the errors are possible systematics, e.g. based on the linelist and uncertainties in the 3D and/or NLTE corrections, or possible offsets in the scale of our adopted methods of deriving \teff{} and \logg. 

\section{Details of the fitting procedure}\label{sec:fittingdetails}

If J1010+2358 inherited a fraction $f_k \in [0,1]$ of its metals from a SN of type $k$ and the remainder $f_j=1-f_k$ from a second SN of type $j$, then its chemical abundance pattern would be described by:
\begin{equation}
\small
     F_i = {\rm [X/Fe]} = {\rm log} \left( \frac{f_k{\rm {\rm Y_Z}^{j} Y_X^{k}}+f_j{\rm Y_Z^{k}}{{\rm Y_X}^{j}}}{f_k{\rm {\rm Y_Z}^{j} Y_{Fe}^{k}}+f_j {\rm Y_Z^{k}}{{\rm Y_{Fe}}^{j}}} \right) \\
    - {\rm log} \left( \frac{\rm M_X}{\rm M_{Fe}} \right)_\odot 
\end{equation}
where ${\rm Y_X}^{k}$, ${\rm Y_X}^{j}$ and ${\rm Y_Z}^{k}$, ${\rm Y_Z}^{j}$ are the theoretical elemental yields and total metal yields of the two SN progenitors, respectively. 

For each pair of SN progenitors (and each $f_k$), we then assess the goodness of fit to the observed abundances, $D_i$, of J1010+2358 using the standard formula \citep{Heger2010}:


\begin{equation}
    \chi^2=\sum_{i=1}^N \frac{(D_i-F_i)^2}{\sigma^2_{i}} + \sum_{i=N+1}^{N+U} \Theta(D_i-F_i) \times \infty + \sum_{i=N+U+1}^{N+U+L} \Theta(F_i-D_i) \times \infty, 
\label{e:chi}  
\end{equation}

where $N, U$ and $L$ are respectively the number of finite measurements, upper and lower limits, $\sigma_i$ are the observational uncertainties (Table~\ref{tab:abu}) and $\Theta(x)$ is the Heaviside function. We treat upper and lower limits very strictly, imposing that the $\chi^2$ of the model goes to infinity if they are not respected.

\end{document}